\documentclass[10pt]{iopart}

\usepackage{graphicx}  
\usepackage{lineno}
\usepackage{cite} 
\begin{document}
	
	
\article[Evidence for the Confinement of Magnetic Monopoles in Quantum Spin Ice]{Letter to the Editor}{Evidence for the Confinement of Magnetic Monopoles in Quantum Spin Ice}

\author{P~M~Sarte$^{1,2}$, A~A~Aczel$^3$, G~Ehlers$^3$, C~Stock$^{2,4}$, B~D~Gaulin$^{5,6,7}$, C~Mauws$^{12}$, M~B~Stone$^3$, S Calder$^3$, S~E~Nagler$^3$, J~W~Hollett$^8$, H D Zhou$^{9}$, J S Gardner$^{10,11}$, J~P~Attfield$^{1,2}$  \& C~R~Wiebe$^{5,6,8,12}$}

\address{$^{1}$ School of Chemistry, University of Edinburgh, Edinburgh EH9~3FJ, United Kingdom} 
\address{$^{2}$ Centre for Science at Extreme Conditions, University of Edinburgh, Edinburgh EH9~3FD, United Kingdom}
\address{$^{3}$ Quantum Condensed Matter Division, Oak Ridge National Laboratory, Oak Ridge, TN 37831, USA}
\address{$^{4}$ School of Physics and Astronomy, University of Edinburgh, Edinburgh EH9~3FD, United Kingdom}
\address{$^{5}$ Department of Physics and Astronomy, McMaster University, Hamilton, ON L8S~4M1, Canada}
\address{$^{6}$ Canadian Institute for Advanced Research, Toronto, ON M5G~1Z8, Canada}
\address{$^{7}$ Brockhouse Institute for Materials Research, McMaster University, Hamilton, Ontario, L8S~4M1, Canada}
\address{$^{8}$ Department of Chemistry, University of Winnipeg, Winnipeg, MB R3B~2E9 Canada}
\address{$^{9}$ Department of Physics and Astronomy, University of Tennessee, Knoxville, Tennessee 37996, USA}
\address{$^{10}$ Center for Condensed Matter Sciences, National Taiwan University, Taipei 10617, Taiwan}
\address{$^{11}$ Neutron Group, National Synchrotron Radiation Research Center, Hsinchu 30076, Taiwan}
\address{$^{12}$ Department of Chemistry, University of Manitoba, Winnipeg, MB R3T~2N2, Canada}
\ead{\mailto{ch.wiebe@uwinnipeg.ca}}
\vspace{10pt}
\begin{indented}
\item[]Received August 2017
\item[]Accepted 
\item[]Published
\end{indented}

\begin{abstract}
Magnetic monopoles are hypothesised elementary particles connected by Dirac strings that behave like infinitely thin solenoids [Dirac 1931 \emph{Proc. Roy. Soc. A} \textbf{133} 60]. Despite decades of searches, free magnetic monopoles and their Dirac strings have eluded experimental detection, although there is substantial evidence for deconfined magnetic monopole quasiparticles in spin ice materials [Castelnovo, Moessner \& Sondhi 2008 \emph{Nature} \textbf{326} 411]. Here we report the detection of a hierarchy of unequally-spaced magnetic excitations \emph{via} high resolution inelastic neutron spectroscopic measurements on the quantum spin ice candidate Pr$_{2}$Sn$_{2}$O$_{7}$. These excitations are well-described by a simple model of monopole pairs bound by a linear potential [Coldea \emph{et al.}~\emph{Science} \textbf{327} 177] with an effective tension of 0.642(8) K~$\cdot$\AA$^{-1}$ at 1.65~K. The success of the linear potential model suggests that these low energy magnetic excitations are direct spectroscopic evidence for the confinement of magnetic monopole quasiparticles in the quantum spin ice candidate Pr$_{2}$Sn$_{2}$O$_{7}$.

\end{abstract}
%
%
%
\maketitle
%
%

\section{Introduction}~\label{sec:introduction} 

Magnetic monopoles remained at the periphery of physics until Dirac published his quantum theory of magnetic charge~\cite{dirac1931} in which he envisioned a monopole as the end of an infinitesimally thin solenoid construct known as a Dirac string. Dirac proposed that not only were magnetic monopoles consistent with quantum theory, but their existence would result in the quantisation of electrical charge~\cite{dirac1931, morrisdirac}. While the latter has been verified experimentally~\cite{oildrop}, the identification of magnetic monopoles has been challenging.~Establishing the existence of this elusive elementary particle would lead to a beautiful symmetrisation of Maxwell's equations and validate several modern physical theories~\cite{search2}.

\begin{figure}
	\begin{center}
		\scalebox{0.333}{\includegraphics{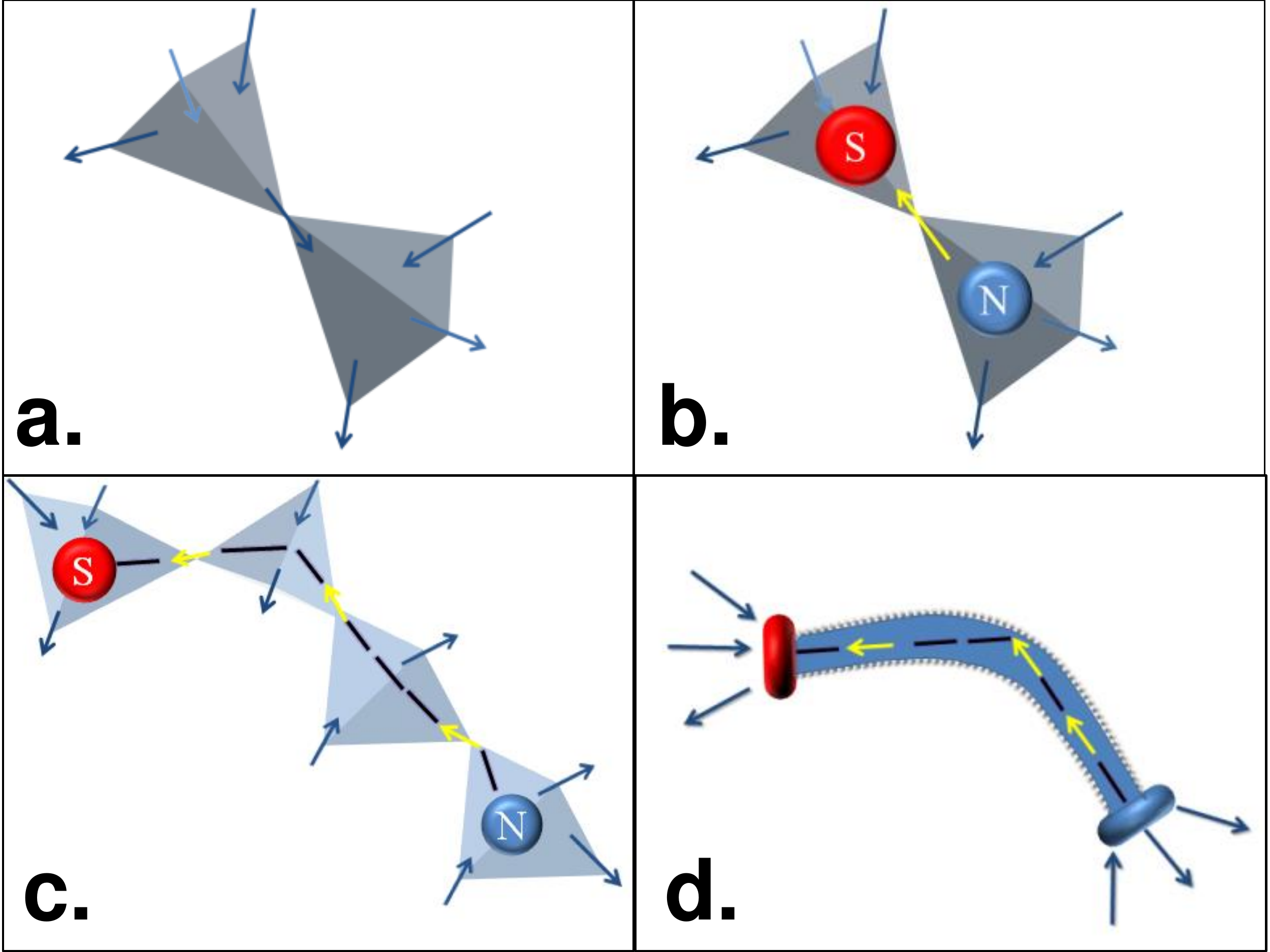}}
		\caption{\textbf{Local spin configurations in quantum spin ices leading to magnetic monopoles and Dirac strings.} (a) Schematic of one possible two-in/two-out spin ice configuration in adjacent tetrahedra of the pyrochlore lattice. (b) A defect spin ice state is created by the flipping of a spin labeled in yellow and results in the creation of a magnetic monopole pair labeled N and S. (c) The monopole pair can separate further via adjacent spin flips. (d) A schematic of an effective ``Dirac string", which consists of an infinitesimally thin solenoid (one unit of flux width) connecting the monopole pair.}
		\label{fig:figure1}
	\end{center}
\end{figure}

Recently, the discovery of a class of magnets known as spin ices has made the study of magnetic monopole quasiparticles viable~\cite{castelnovo, morrisdirac, bramwell}. Spin ices are found in a series of magnetic pyrochlore oxides A$^{3+}_{2}$B$^{4+}_{2}$O$_{7}$, which have moments residing on the A-site, corner-sharing tetrahedra sublattice. At low temperatures, the moments assume a two-in/two-out short-ranged magnetically ordered state as shown in Figure~\ref{fig:figure1}(a), possessing Pauling's configurational entropy~\cite{spiniceramirez}. Castelnovo \emph{et al.}~\cite{castelnovo} first proposed that dipolar spin ices (DSIs) may host mobile mangetic monopole quasiparticles as illustrated in Figure~\ref{fig:figure1}(b). These monopoles are expected to interact \emph{via} a magnetic Coulomb law suggesting deconfinement~\cite{castelnovo, morrisdirac}, and the strings connecting them in pairs (see Figures~\ref{fig:figure1}(c) and~\ref{fig:figure1}(d)) have not been easily measurable. Consequently, although there is mounting experimental evidence~\cite{morrisdirac, diracstring} supporting the existence of monopoles in the DSIs, the exact nature of the interaction between these monopoles is still under active investigation\cite{paulsen}.

In an attempt to measure interactions between magnetic monopoles, our attention has shifted to quantum spin ices (QSIs)~\cite{gingrasspinice,sibille,wen,QSI_more,light}. This family of materials differs from DSIs in the nature of the interactions between the magnetic moments~\cite{gingrasspinice}, as their magnetic Hamiltonians consist of transverse coupling terms leading to significant fluctuations of the moments away from the local [111] quantisation axes. As a result, the correlation time of the two-in/two-out state at low temperatures tends to be much shorter for a QSI as compared to its DSI counterparts~\cite{jason}. There are predictions~\cite{negative2} for the properties of monopoles in QSIs, but their detection has remained elusive. We report here the direct observation of interacting magnetic monopoles in Pr$_{2}$Sn$_{2}$O$_{7}$ using inelastic neutron spectroscopy. Our measurements allow for both an estimate of the monopole pair creation energy and a lower bound of the effective tension between monopoles.

\section{Experimental}~\label{experimental} 

\begin{figure*}
	\centering
	\scalebox{0.43}{\includegraphics{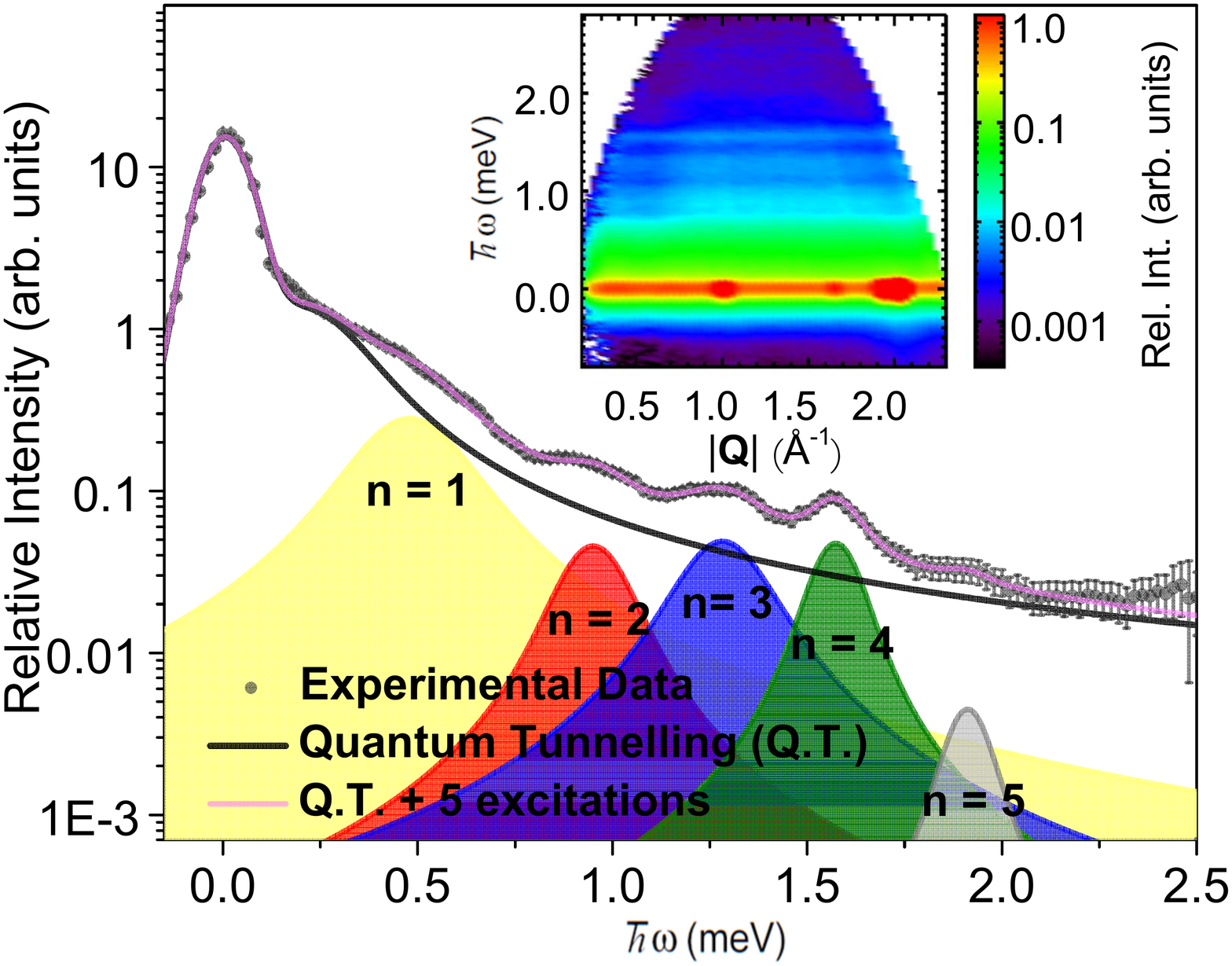}}
	\caption{\textbf{Fine structure in the low energy excitation spectrum of Pr$_{2}$Sn$_{2}$O$_{7}$.} A $|\mathbf{Q}|$-integrated ($\rm{|\mathbf{Q}| = [1.0, 1.5]~\AA^{-1}}$) cut along $\hbar\omega$ of (inset) the empty aluminium can-subtracted S($|\mathbf{Q}|, \omega$) spectrum of polycrystalline Pr$_{2}$Sn$_{2}$O$_{7}$ at 1.65 K with an incident energy $\hbar\omega_{i}$~=~3.32~meV on CNCS operating in high flux mode, displaying unequally-spaced nearly-dispersionless excitations. The high flux mode was accomplished by spinning the high speed double disk chopper located just before the sample at a frequency of 300~Hz. A fitting routine, as described in~\emph{Supplementary Discussion \S 3.3.}, was applied exclusively to positive energy transfers with the fit to each excitation shown and labelled by its quantum number $n$. The strong, broad Lorentzian feature centered at approximately 0.2~meV was observed in previous work~\cite{zhou2008} and is attributed to quantum tunnelling between degenerate spin ice configurations. The relative intensity axis is presented on a logarithmic scale for both the constant $|\mathbf{Q}|$-cut and S($|\mathbf{Q}|, \omega$) spectrum inset to assist with data visualisation due to the relatively low intensity of the excitations as compared to the elastic line.}  
	\label{fig:figure2}
\end{figure*}

Polycrystalline samples of Pr$_{2}$Sn$_{2}$O$_{7}$~were prepared by a standard solid-state reaction of stoichiometric amounts of Pr$_{6}$O$_{11}$ (99.99~\%, Alfa Aesar) and SnO$_{2}$ (99.99~\%, Alfa Aesar). The powder reagents were mixed together, finely ground and pressed into a pellet using a uniaxial press. The pellets were placed in an alumina crucible and were pre-reacted by heating in air at 1000$\rm{^{\circ}}$C for 24~h. The pellets were then reground, repelletised and heated in air at 1400$\rm{^{\circ}}$C for approximately 48~h with intermittent grindings until room temperature powder x-ray diffraction measurements with a Bruker D2 phaser laboratory diffractometer at the University of Edinburgh using a Cu~K$_{\alpha_{1,2}}$ source confirmed no discernable impurities.

Low energy transfer inelastic neutron scattering experiments were performed on the direct-geometry time-of-flight cold neutron chopper spectrometer CNCS at the Spallation Neutron Source (SNS) at Oak Ridge National Laboratory (ORNL). Approximately five grams of polycrystalline Pr$_{2}$Sn$_{2}$O$_{7}$ and select members of Pr$_{2}$Sn$_{2-x}$Ti$_{x}$O$_{7}$ ($x$ =~0.40 and 0.60) were sealed in aluminium cans under a helium atmosphere for the experiment. The sample cans were mounted on the CNCS automatic three sample rotator stick (SS-003) with a boron nitride (BN) spacer adapted for a top loading 100~mm orange cryostat (CRYO-006). Measurements utilised incident energies $\hbar \omega_{i}$ of 3.32~meV and 25~meV in high flux mode, providing an energy resolution at the elastic line of approximately 0.01 and 2~meV, respectively. Additional measurements were collected with an incident energy of 4.1~meV in medium resolution mode, providing an energy resolution at the elastic line of approximately 0.08~meV. An empty aluminium can was also measured for approximately half the counting time at identical experimental conditions, and the resulting spectra were subtracted from the corresponding sample spectra. The high flux and medium resolution modes were accomplished by spinning the high speed double disk chopper located just before the sample at a frequency of 300~Hz and 240~Hz, respectively.

\section{Results \& Discussion}~\label{sec:results}

The pyrochlore Pr$_{2}$Sn$_{2}$O$_{7}$ has been well characterised as a potential QSI candidate~\cite{zhou2008, gingrasspinice}. Despite susceptibility measurements suggesting net ferromagnetic interactions ($\rm{\theta_{CW}}$~=~0.3~K), there is an absence of long-range magnetic order~\cite{lowPr2Sn2O7,zhou2008}. The ground state crystal field scheme is well understood~\cite{zhou2008, princep2}, consisting of a thermally-isolated non-Kramers doublet. The lower Pr$^{3+}$ effective moment of 2.61(1)~$\rm{\mu_{B}}$, as compared to $\sim$10~$\rm{\mu_{B}}$ for DSIs, implies Pr$_{2}$Sn$_{2}$O$_{7}$ is more susceptible to quantum fluctuations~\cite{princep2}. In fact, low energy spin fluctuations persist to well below 1~K and possess an anomalously low activation energy, which is attributed to the quantum nature of the system\cite{zhou2008}.

\begin{figure}[h!]
	\centering
	\scalebox{1}{\includegraphics{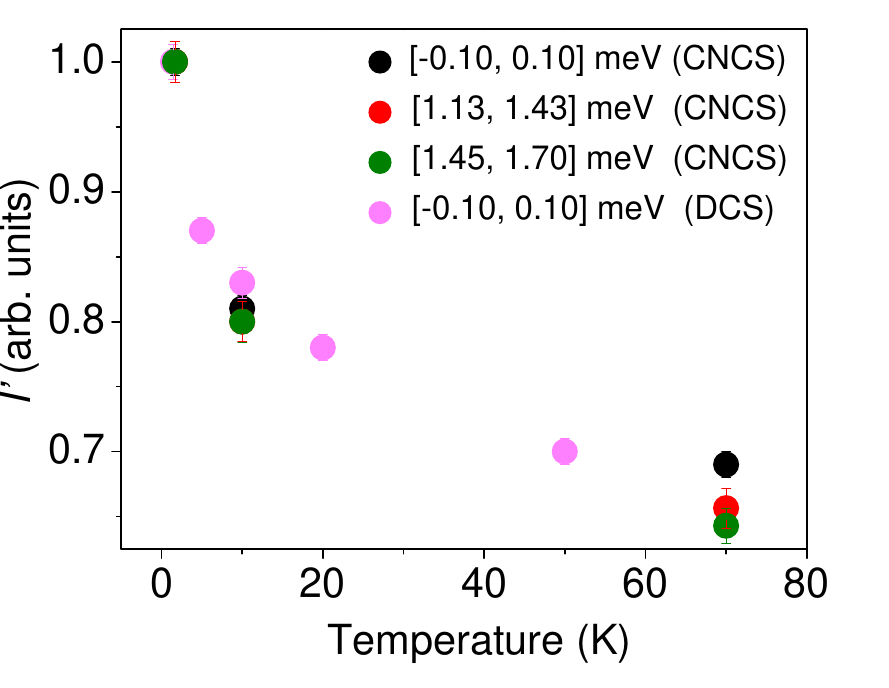}}
	\caption{\textbf{Temperature dependence of the magnetic scattering in Pr$_{2}$Sn$_{2}$O$_{7}$.} Comparison of the temperature dependence of the normalised integrated intensity $I'$ of the magnetic diffuse scattering and the two lowest energy magnetic excitations observed at CNCS. For the purposes of comparison, the temperature dependence of the integrated intensity of the magnetic diffuse scattering from previous DCS data~\cite{zhou2008} is also shown. Both data sets are integrated over the same range in $|\mathbf{Q}|$~=[0.35,1.75]~\AA$^{-1}$. The similar temperature dependence of all four data sets suggests that the low energy magnetic excitations are associated with the quantum spin ice state. Please refer to \emph{Supplementary Discussion} \S 3.4. for further details.}  
	\label{fig:figure3}
\end{figure}

We have remeasured these low-energy spin fluctuations in a well-characterised powder sample of Pr$_{2}$Sn$_{2}$O$_{7}$ using the Cold Neutron Chopper Spectrometer (CNCS). Broad, quasi-elastic scattering previously measured~\cite{zhou2008} was confirmed in our sample at 1.65~K. However, the high resolution and time-integrated flux of the CNCS also enabled the observation of a discernible fine structure to the scattering, as shown in Figure~\ref{fig:figure2}. A hierarchy of nearly-dispersionless excitations was identified and measured up to $\hbar\omega$~$\sim$ 2~meV. These excitations are also visible with different incident energies while obeying detailed balance, confirming they are not spurious in origin. Furthermore, as shown in Figure~\ref{fig:figure3}, they exhibit a remarkably similar temperature dependence to the magnetic diffuse scattering observed in the elastic channel. This similar temperature dependence suggests that the excitations are associated with the quantum spin ice state. Finally, these modes are not due to chemical disorder, as this is well understood through doping studies of Pr$_{2}$Sn$_{2-x}$Ti$_{x}$O$_{7}$ and related materials~\cite{gaulin}.~Please refer to \emph{Supplementary Discussion} \S 3 for further details. 

As summarised by Figure~\ref{fig:figure2}, the excitation spectrum was analysed by fitting the data to a series of free-width Lorentzians convolved with fixed-width Gaussians. The lowest-energy Lorentzian function models a prominent quasi-elastic background centered at 0.2~meV, previously attributed to quantum tunnelling between degenerate spin ice configurations~\cite{zhou2008}. The higher-energy Lorentzians represent the five quantised excitations observed here. The analysis reveals that these modes are not evenly-spaced and decrease in relative intensity with increasing temperature, which taken together rule out a possible quantum harmonic oscillator interpretation~\cite{aczel} and imply a magnetic origin instead. We therefore consider possible mechanisms that lead to quantised, unevenly-spaced magnetic excitations. The dynamic response for the spin defects of a quasi-one-dimensional Ising $S$~$=$~$\frac{1}{2}$ spin chain such as CoNb$_2$O$_6$~\cite{coldea} below the magnetic ordering temperature has the desired characteristics. While the spin defects are created in pairs, they are ultimately bound together in the ordered state due to an attractive linear potential arising from the finite molecular field~\cite{torrance1,torrance2,shiba}. A similar spin defect confinement model may apply to Pr$_{2}$Sn$_{2}$O$_{7}$, under the assumption that the relevant defects in this case are magnetic monopole quasiparticles and not solitons, while the linear confining potential may be a consequence of the QSI state. Note that the analogy between spin defects and monopoles in a spin ice has been discussed previously~\cite{bramwell, morrisdirac}. If the monopole confinement model is valid for Pr$_{2}$Sn$_{2}$O$_{7}$, then the energies of the excitations should be described by the following:
	\begin{equation}
	\hbar\omega_{n} = 2\hbar\omega_{\rm o} + z_{n}\lambda^{\frac{2}{3}}\left(\frac{\hbar^{2}}{\mu} \right)^{\frac{1}{3}},  
	\label{eq:solution}
	\label{eq:1}
	\end{equation}
where $n$ is a positive integer, $\mu$ is the reduced mass, $\hbar\omega_{n}$ is the $n^{\rm th}$ excitation energy, 2$\hbar\omega_{\rm o}$ is the energy cost to produce a pair of monopoles, $\lambda$ is an effective tension and $z_{n}$ are the negative zeros of the Airy function~\cite{mccoy, coldea}. Furthermore, if one fixes $\mu$ to the appropriate value based on previous work by Pan~\emph{et al.}~\cite{pan} on another quantum spin ice candidate Yb$_{2}$Ti$_{2}$O$_{7}$, then the linear relationship between $\hbar\omega_{n}$ and $z_{n}$ also provides an estimate for the lower bound of the effective tension $\lambda$ between monopoles \emph{via} the slope of Equation~\ref{eq:1}. Please refer to \emph{Supplementary Discussion \S 3.3-3.5} for further details. 

\begin{figure}
	\centering
	\scalebox{1}{\includegraphics{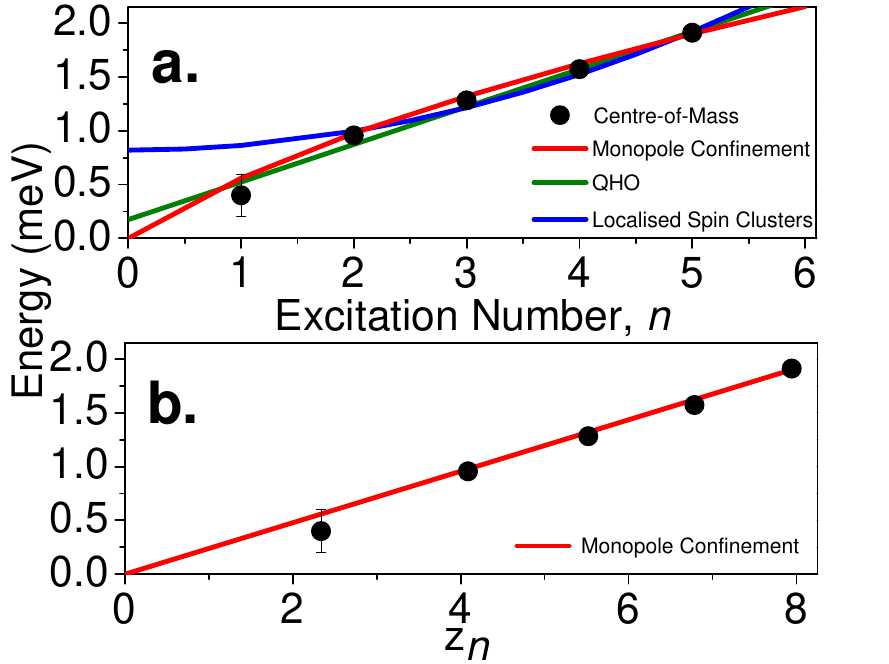}}
	\caption{\textbf{Validation of the monopole confinement model for Pr$_{2}$Sn$_{2}$O$_{7}$.} (a) The direct comparison between the observed low energy magnetic excitations at 1.65~K and the predicted behaviour from three competing models: quantum harmonic oscillator~\cite{aczel}, localised spin clusters~\cite{stock}, and monopole confinement~\cite{coldea}. The predicted values of $\hbar\omega$ for each model were obtained by fitting the corresponding observed values to each model's fitting function as described in the main text, and are connected by interpolated lines as a guide to the eye.~The calculated goodness-of-fit metric $\chi_{\nu}^{2}$ is the lowest for the monopole confinement model. (b) The plot of excitation energies against the negative zeros of the Airy function $z_{n}$ exhibits linear behaviour as predicted by the monopole confinement model~\cite{coldea}.~A least squares fit with the linear Equation~\ref{eq:1} in the main text (shown by the solid red line in both panels (a) and (b)) yielded a lower bound for the effective tension $\lambda$ of 0.642(8)~K~$\cdot$~\AA$^{-1}$ and a monopole pair creation energy 2$\hbar\omega_{\rm o}$ of 0~meV at 1.65~K (see \emph{Supplementary Discussion \S 3.5.2. and~\S 3.5.3.}).}
	\label{fig:figure4}  
\end{figure}

Figure~\ref{fig:figure4}(a) plots the observed energy levels versus the excitation number to facilitate a direct comparison between three different candidate scenarios. Two possibilities, the monopole confinement and QHO~\cite{aczel} models, have already been discussed in detail above. We consider a third model here, based on localised high $S$ clusters. The magnetic excitation spectrum for an isolated spin cluster also consists of a series of quantised energy levels, and the spacing between the modes can be non-trivial or follow the simple relationship $\hbar\omega_{n}$~$\propto$~$n^2$ depending on the specific details of the magnetic Hamiltonian~\cite{16_rau,stock}. We fit our data to each of these models with the phenomenological expression $\hbar\omega_{n} = A + B x_n$, where $A$ and $B$ are constants and $x_n$ is $n$, $n^2$, or $z_n$ for the QHO, localised spin cluster or monopole confinement models, respectively. The solid curves in Figure~\ref{fig:figure4}(a) represent the best fit to each of the models and they clearly illustrate that the monopole confinement model provides the best agreement. Additional details are presented in Table~\ref{tab:1}. Furthermore, a plot of $\hbar\omega_{n}$ vs.~$z_{n}$, as shown in Figure~\ref{fig:figure4}(b), produces a linear relationship in agreement with the predictions of this model~\cite{coldea}. Therefore, we interpret these excitations as direct spectroscopic evidence of interacting magnetic monopoles within the pyrochlore lattice. 

The linear fit shown in Figure~\ref{fig:figure4}(b) yields zero as an estimate for 2$\hbar\omega_{\rm o}$ and a lower bound of 0.642(8)~K~$\cdot$~\AA$^{-1}$~for the effective tension between monopoles at 1.65~K. Some physical insights can be made from the measurement of these two parameters. Firstly, the value of the tension is positive and non-negligible, implying that the monopoles are confined unlike in DSIs~\cite{morrisdirac, bramwell}. Secondly, the application of an identical analysis algorithm to data collected at 10~K yields a larger lower bound of 0.667(8)~K~$\cdot$~\AA$^{-1}$~for the effective tension, an increase that would be expected if the confining potential was attributed --- to some extent --- to the spin ice state. Thirdly, the value of $\lambda \sim$~0.6~K$\cdot~\rm{\AA^{-1}}$ is surprisingly large for the expected energy scale with $J \sim 1$~K~\cite{zhou2008}.~In fact, the energy cost to separate two monopoles by the distance between the centres of adjacent Pr$^{3+}$ tetrahedra corresponds to approximately a temperature scale of 3~K. This strong tension prevents the propagation of monopoles over long distances. The confinement of these monopoles can be roughly quantified, since exact analytical solutions for the Schr\"{o}dinger equation with an $|x|$-potential are known~\cite{hohlfeld}. The expectation value $\langle |x| \rangle$ for the highest energy excitation clearly observed, $n$~$=$~5, corresponds to a relatively short distance of approximately 20~\AA, or two unit cells. Finally, it should be noted that although the linear fit of both 1.65~K and 10~K data yields a value of zero (within error) as an estimate of 2$\hbar\omega_{\rm o}$, its absolute value is extremely sensitive to the energy of the first excitation and thus should be interpreted with caution. Please refer to \emph{Supplementary Discussion \S 3.5.} for further details. 

	\begin{table*}
	\caption{Comparison between the observed and predicted values of the energy transfer $\hbar\omega$ for the quantum harmonic oscillator, localised spin cluster and monopole confinement models at 1.65~K. The calculated values were obtained by fitting the five observed low energy excitations to the functional form of the individual models, with further details of the fitting described in the main text. Numbers in parentheses indicate statistical errors. The goodness-of-fit metric $\chi_{\nu}^{2}$ (see \emph{Supplementary Discussion \S 3.3.}) for each model is provided to enable quantitative comparison.~\label{tab:1}}
	\hspace*{10.00mm} 
	\begin{tabular}{@{}llllll}
		\br
		Model $\biggl/$ Mode & $n$ = 1 & $n$ = 2 & $n$ = 3  & $n$ = 4  & $n$ = 5\\
		\mr
		Observed (1.65~K)  &  0.4(2) &  0.96(2) &  1.28(4)   &  1.57(2)  &  1.912(9)  \\ [1ex] \hline
		Monopole Confinement ($\chi_{\nu}^2$~$=$~5.7) &  0.559(5) & 0.978(8) & 1.32(1) & 1.62(1) & 1.90(2)  \\ [1ex] \hline 
		Quantum Harmonic Oscillator ($\chi_{\nu}^2$~$=$~22) & 0.524(5) &  0.873(9)  & 1.22(1)  & 1.57(2)  & 1.92(2) \\ [1ex] \hline 
		Localised Spin Clusters ($\chi_{\nu}^2$~$=$~8.6) & 0.86(6) &  1.00(7) & 1.22(9) & 1.5(1)  & 1.9(1) \\
		\br
	\end{tabular}
\end{table*}

\section{Conclusion}~\label{sec:conclusion} 

High resolution cold inelastic neutron scattering measurements on polycrystalline Pr$_{2}$Sn$_{2}$O$_{7}$ have revealed a previously unreported fine structure to the low energy excitation spectrum consisting of a series of unevenly spaced nearly-dispersionless magnetic excitations. A quantum confinement model with a linear potential $\lambda|x|$ accounts for the fine structure suggesting these magnetic excitations are a direct spectroscopic observation of interacting magnetic monopole quasiparticles resulting from a finite tension between them. One natural extension of this work would be to remeasure the low energy dynamics of other QSI candidates (e.g. Pr$_{2}$Zr$_{2}$O$_{7}$) to determine if these systems exhibit similar non-negligible monopole tensions, while another prospect is to explore the effects of external perturbations on the tension such as the application of external magnetic fields and pressure. The success of the monopole confinement model for Pr$_{2}$Sn$_{2}$O$_{7}$ encourages future QSI studies with the ultimate goal of understanding exactly how monopole confinement affects other physical properties of QSIs.

\ack{This work was supported by NSERC, ACS, PRF, CFI, NSF, STFC and EPSRC. PMS acknowledges the University of Edinburgh for funding through the Global Research and Principal's Career Development Scholarships and the Canadian Centennial Scholarship Fund. CRW acknowledges financial support from the CRC (Tier II) program, PRF, CIFAR and NSERC. JWH and CM acknowledges financial support from NSERC. JPA and CS acknowledges financial support from the STFC, EPSRC and the ERC. Concerning previously obtained data from the DCS, CRW and HDZ acknowledge financial support from the NSF, the EIEG program (FSU) and the state of Florida. We acknowledge the support of the National Institute of Standards and Technology, U.S. Department of Commerce, in providing the neutron research facilities used in this work. A portion of this research at Oak Ridge National Laboratory's Spallation Neutron Source was sponsored by the Scientific User Facilities Division, Office of Basic Energy Sciences, U.S. Department of Energy.  Finally, the authors thank the Carnegie Trust for the Universities of Scotland for providing facilities and equipment for chemical synthesis.}

\section*{References}

\bibliographystyle{iopart-num}
\bibliography{references}

\end{document}